\documentclass[aip,sd,preprint,graphicx]{revtex4-1} 

\usepackage{amsmath,amssymb,eucal,graphicx,subfigure}
\usepackage{color}
\begin{document}

\title{A comparison of the innate flexibilities of six chains in F$_1$-ATPase  with identical
secondary and tertiary folds; 3 active enzymes and 3 structural proteins}

\author{Monique M. Tirion}
\email{mmtirion@clarkson.edu}

\affiliation{Physics Department, Clarkson University, Potsdam, New York
13699-5820, USA} 

\begin{abstract}


The $\alpha$ and $\beta$
 subunits comprising the hexameric assembly of F1-ATPase share a high degree of structural identity, though low primary  identity.  Each subunit  binds  nucleotide in similar  pockets, yet  only $\beta$ subunits are  catalytically active. Why?  We re-examine their internal symmetry axes and observe  interesting differences. Dividing each chain into an N-terminal head region, a C-terminal foot region, and a central torso, we observe (1) that while the foot and head regions in all chains obtain high and similar mobility,  the torsos obtain different mobility profiles, with the $\beta$ subunits exhibiting a  higher  motility compared to the $\alpha$ subunits, 
a trend supported by the crystallographic B-factors.  The $\beta$ subunits have greater torso mobility by having fewer distributed, nonlocal packing interactions providing a spacious and soft connectivity, and offsetting the resultant softness with local stiffness 
elements, including an additional  $\beta$ sheet. 
(2) A loop near the nucleotide binding-domain of the $\beta$ subunits, absent in the $\alpha$ subunits, 
swings to create a large variation in the occlusion of the nucleotide binding region. (3) A combination of the softest three eigenmodes significantly reduces the RMSD between  the open and closed conformations of the $\beta$ subnits. 
(4) Comparisons of computed and observed crystallographic B-factors suggest a suppression of a particular 
symmetry axis in an $\alpha$ subunit. 
(5) Unexpectedly, the soft intra-monomer  oscillations  pertain to  distortions that do not create inter-monomer steric 
clashes in the assembly, suggesting that  structural optimization of the assembly evolved at all levels of complexity.

\end{abstract}

\maketitle

\section{INTRODUCTION}
\subsection{Overview of hexameric F$_1$-ATPase}

ATP synthases exploit ion gradients generated during electron transport reactions at cell interfaces to phosphorylate ADP  
and replenish the cell's supply of ATP.  Mild salt treatments dissociate ATP synthases into  two
fractions: a membrane-embedded F$_o$ portion and a soluble, hydrophilic F$_1$ portion 
(for reviews, see \cite{Xu,WalkerRev,senior}).  In the intact enzyme, the F$_o$ portion
links an ionic gradient to a mechanical rotation, while the F$_1$ portion channels the rotary motion to the  
synthesis reaction.   The dissociated F$_1$ portion lacks the capacity to generate ATP, however it does function as
an ATPase, hydrolyzing ATP in the presence of ATP, ADP and phosphate, P$_i$.
The isolated F$_1$  complex consists
of five different protein chains with stoichiometries of $\alpha_3\beta_3\gamma_1\delta_1\epsilon_1$
and mass ratios 55, 51, 31, 15 and 6 kD.
F$_1$ fractions obtained from bacteria, fungi, animals and plants have been crystallized and their atomic positions resolved and
reveal this enzyme complex to possess a highly conserved structure \cite{Nelson,Yoshida}.

\begin{figure}[h]
\includegraphics[width=0.4\textwidth]{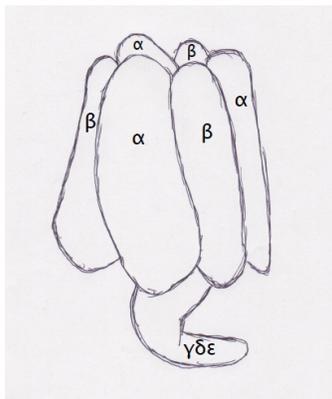}
\caption{Schematic of the F$_1$-ATPase fragment of ATP synthase. Composed of alternating $\alpha$ and $\beta$
subunits, the central axis of pseudosymmetry obtains an  $\alpha$-helical coiled coil from the N and C terminal
domains of the $\gamma$ subunit, while the remainder of that subunit plus the $\delta$ and $\epsilon$ chains
protrude from the central channel below the foot domains.  The hexameric cap has a diameter of around
100\AA.}
\label{Figure01}
\end{figure}

As illustrated in Figure \ref{Figure01},
in F$_1$-ATPase   the three $\alpha$ subunits (SUA) and  three $\beta$ subunits (SUB) 
alternate as the segments of an orange to create a cap-like
structure with an outer diameter of around 100\AA~ and a central channel about 20\AA~ across.
This central channel, marking the axis of pseudosymmetry,
contains a pair of coiled-coil $\alpha$ helices formed by the N and C terminal domains of the $\gamma$ subunit.
The remainder of the $\gamma$ chain as well as the smaller $\delta$ and $\epsilon$ chains form a globular arrangement
attached to the central $\alpha$ helices like the head of a golf club to its shaft.

The X-ray structures show  the $\alpha$ and $\beta$ chains to possess nearly identical three-dimensional conformations
with all-atom root mean square difference (RMSD) superpositions  between 2.2 and 2.6\AA, but with 
primary sequence identity and similarity of  25\% and 43\% \cite{needleman}.
Adenosyl nucleotides can bind to each SUA and SUB in binding pockets located at their interfaces.
However, only SUB  is catalytically active: 
ATP bound to SUA  is neither hydrolyzed nor exchanged with solvent medium \cite{Boyer89,Senior90,malyan}.  
Catalysis at the three $\beta$ subunits occurs not with use of high energy intermediates, 
but in a cooperative, cyclic fashion termed the binding change mechanism  \cite{BoyerRev}.
Studying heavy oxygen exchange rates during ATP synthase catalysis in the presence and absence of a proton gradient,
Boyer realized  that
the $pmf$ at F$_o$ is energetically coupled with product release at F$_1$
rather than  chemical bond-formation. Once bound to a catalytic site, in other words, ADP and P$_i$ 
 spontaneously interconvert to ATP  without external energy and   have an equilibrium constant close to 1.  
According to the binding change mechanism, each $\beta$ subunit sequentially
binds ADP and P$_i$, then undergoes a conformational transition and makes ATP, and finally changes conformation
again with release of product. The three subunits function  in concert, with each subunit cycling through the same
states consecutively, so that the system``hangs" if one subunit is prevented from transitioning, by for example,
removal of product  from  solvent medium.  Experiments suggest 
that the rotary, cyclic behavior of the hexamer persists during hydrolysis in the absence
of the $\gamma$ chain, though with lower precision and rate constants  \cite{Miwa,Noji,Junge}.

Our current analyses will focus on the elements
comprising this minimal functional unit, the $\alpha$ and $\beta$ chains.
In particular,  we examine the question:
why do $\beta$ subunits readily hydrolyze ATP and exchange the HOH generated with medium water,
while the $\alpha$ subunits neither hydrolyze nor exchange ATP with solvent nucleotides? 
Xu and coworkers \cite{Xu} point out that while the nucleotide-binding sites in $\alpha$ and $\beta$ subunits are closely conserved, one carboxylate, 
of residue $\beta$-Glu 188
is replaced by $\alpha$-Gln 208,  eliminating a likely catalytic base in the $\alpha$ subunits. 
Furthermore, Xu points out that the $\alpha$ subunit's ``inability to transition
between different catalytic conformations ... as evidenced by the absence of open conformation" in crystalline structures,
severely dampens  their catalytic activity.
In this work we closely examine the extent and reason for the ``inablity'' of $\alpha$ subunits to cycle through the
conformations adopted by the $\beta$ subunits. Superficially, one might expect two proteins with such
similar fold and architecture to exhibit similar flexibility characteristics. In particular, then, the internal
symmetry axes of these different protein chains should be nearly identical. Are they?  We compute and examine
each proteins' slowest eigenmodes via PDB-NMA. 
We observe interesting similarities
and differences that may help further explain the different catalytic propensities of these two subunits.

\subsection{Overview of PDB-NMA}

Just as  rigid structures obtain 3 rotational  principal symmetry axes determined by
diagonalization of their inertia matrix,   nonrigid objects obtain internal
symmetry axes derived from diagonalization of their Hessian matrix \cite{Marion}. An object has
 as many internal symmetry axes or eigenvectors as it has internal degrees of freedom, with each
eigenvector a specific linear combinatiton of those internal degrees of freedom.
The Hessian matrix of an object describes the distribution, not of masses, but of forces about a position of stability
as its internal degrees of freedom are varied.
 Rigid principal
axes pertain to spatial symmetries where the mass distributions about each of  3 special, orthogonal
axes are balanced  and the  angular velocity and momentum vectors aligned.
Nonrigid, internal axes pertain to temporal symmetries: 
each axis pertains to a balanced, internal oscillation of the object with one particular frequency and energy.
These axes are orthogonal and hence  also known as normal modes:  excitation of one mode cannot excite another mode, 
though in practice  anharmonicities introduce off-diagonal elements that decohere motion. 
As the normal modes describe  oscillations innate or intrinsic to the system, the
modes are often referred to as eigenmodes and they form a complete, orthonormal coordinate system.

The set of frequencies obtained by diagonalization of the object's Hessian, called the eigenspectrum,
has a characteristic distribution for proteins,  different from other solids \cite{na,Dani1,Elber}. 
The highest frequencies arise from
 rapid vibrations of the stiffest elements while the slowest frequency is a function
of the mass and shape of the molecule, and is roughly 0.1 THz for a 50kDa protein (spectroscopists attempt to probe such motions
by delivering photons with identical frequencies and often report this in terms of the corresponding photon wavenumber of 
3 $cm^{-1}$.) 

In addition to the eigenfrequencies,
the diagonalization of the Hessian matrix provides the set of vectors or eigenmodes that describe the shape of the motion
associated with each eigenfrequency.
As  folded, globular proteins
obtain  high packing densities and heat capacities comparible to solid crystalline objects \cite{cooper},
it is interesting to ascertain how correlated motions extending over the entire molecule are enabled.
Snapshots of single conformations cannot convey this information and one cannot anticipate how a particular molecule
``solves'' the problem of enabling large-scale motions where thousands (internal) degrees of freedom cooperatively
deform tens of thousands of nonbonded interactions to achieve correlated motions across the entire molecule or
``full body motion,''  FBM.

The innate $dof$s associated with equilibrium oscillations describe motilities
of interest:  energy imparted to an object at rest will be dissipated by these internal degrees of freedom.
Unlike spectroscopists who deliver very precise energy pulses that may
match particular molecular frequencies \cite{hoffmann}, 
thermal baths activate folded protein's  many modes equally.  No particular $dof$
dissipates the thermal energy,  instead the heat energy is distributed amongst all the internal
symmetry axes. The high frequency modes dissipate the kT thermal units in high frequency, 
low amplitude oscillations
while the slowest modes dissipate kT energy units in low frequency, larger amplitude oscillations. There
is a longstanding assumption that
nature exploits  the long-correlation length ``slow'' motions intrinsic in folded proteins to achieve functionality \cite{sanejouand}. 

\subsection{Range of validity}

Normal modes provide another ``quality" or signature of an object much as do mass or charge distributions, 
heat capacity, reflectivity, shape, etc, 
independent of the basis vectors chosen to
describe that object's internal degrees of freedom \cite{Pentland}. 
For this reason, the modes computed using
all-atom Cartesian degrees of freedom or heavy-atom dihedral degrees of freedom or reduced
coordinates such as a single point per residue,  match  for sufficiently slow frequency modes.

All-atom analyses on PDB entries using detailed force fields such as Gromos or Charmm are accurate but handicapped by the
fact that  PDB entries are not at equilibrium according to these parameterizations \cite{levitt1}.  Therefore, prior to diagonalization
of the Hessian, an initial structure-distorting
energy minimization {\it must} be performed. Minimization of objects with thousands of degrees of freedom and tens or hundreds
of thousands of energy terms is not an exact, analytic process, and no single, unique minimized structure exists, even
under a particular force field. Furthermore,  the tens of thousands of nonbonded terms have  exquisitely sensitive
dependencies on their distances of separation 
that result in the rapid accumulation of round-off errors  for the floating point representations
of the energy per conformation. In practice this means that a minimal energy conformation can not be discerned using
double precision computations for proteins larger than around 150 residues. The resulting negative eigenvalues
describe unstable motions that are typically ignored and that complicate the analysis of the remaining positive, stable eigenmodes.
One can avoid these limitations by accepting each PDB entry as representing a stable conformation, a not unreasonable
assumption given that typically $10^{17}$ molecules align identically to provide high resolution X-ray diffraction data 
\cite{ringe}.
This  assumption, that the PDB entry represents a stable, long-lived conformation, 
permits  design of simplified, Hookean force fields to describe equilibrium oscillations \cite{tirion96}.
While initial parameterizations of such Hookean force fields were necessarily
maximally simple, current formulations carefully parameterize every bonded and  nonbonded term 
in accordance to a ``parent'' potential such as ENCAD or CHARMM \cite{song,tirion14}. 
The resultant eigenspectra and eigenmodes
reproduce those obtained using the parent potentials {\it when done on the same, energy minimized coordinates}.  Consequently,
use of these Hookean force fields permits comparison of
the eigenspectra and eigenmodes of two closely similar structures, such as 
the A and B populations within a single protein crystal \cite{tirion15} or
the $\alpha$ and $\beta$  chains of F$_1$-ATPase.

Our current analyses are based on a reduced heavy-atom representation, inclusive all atoms in the PDB entries,
as provided by the force field developed by Michael Levitt, termed L79 \cite{L79}, the precursor to ENCAD \cite{ENCAD}.  
In L79, the energy of each pair of ($\phi,\psi$) mainchain torsions is modeled as a sum of Gaussian-like potentials,
a functional that  reproduces Ramachandran maps and  compensates for the exclusion of nonbonded interactions between atoms separated by three bonds;
while the energy associated with sidechain torsions is modeled by
a  harmonic functional. Nonbonded, dispersive and repulsive forces between all atoms
more than three bond lengths and less than some cutoff distance apart are modeled by a Lennard-Jones 6-12 potential. 
L79 includes only polar hydrogens, with hydrogens bonded to carbons combined into united forms.

This early
formulation of an intra-molecular force field works well for modeling the motility of large, isolated globular proteins.
To better characterize surface forces involved in ligand binding, inter-molecular interactions and water, especially
to reaction pathways, more sophisticated potentials evolved.  Our current interest pertains to analyses of 
entire molecule, full-body motions, FBM, that examines the concurrent activation of thousands of internal $dof$s that cooperatively
deform tens of thousands of NBI.  
While a Hookean force field has a seemingly trivial complexity and range of validity (namely, at a point of equilibrium), 
the space where several thousand constrained degrees of freedom (dihedral angles) exist in a force field
consisting of tens of thousands of
nonbonded pairwise interactions is rich and surprising. Achievement of self-consistency in this vast space, both
in terms of the mathematical formulations as well as in machine encoding, is essential to link effects and causes.
The addition of complexity to this formulation, by inclusion of
all hydrogen atoms and crystal and solvent waters for example, is desirable, but their contributions to the 
FBM can only be discerned and ultimately understood
in contrast to the formulation lacking these elements. 
For this reason, we  characterize the equilibrium motility
spectra of PDB entries primarily using the simpler L79 potential, and only briefly consider the contributions 
within the current formulation of ligands, including nucleotides, cations and crystal HOH.

Normal mode studies of the  F$_1$-ATPase structure with the PDB designation \cite{PDB}
1BMF \cite{Walker1} have been published by Cui and coworkers \cite{cui} 
as well as by Zheng \cite{zheng}.
The latter uses a coarse-grained elastic network model  to study the coupling of 
cyclic conformational transitions,
as modeled by intramonomer hinges and intermonomer rigidy body motions,
and $\gamma$ subunit rotations to ATP binding and product release.
The former, earlier,  study 
used all heavy atoms as well as polar hydrogen atoms,  ligands and crystal waters  in a classical (non-Hookean) force field 
in order to characterize the structural plasticity of the isolated $\alpha$, $\beta$ and $\gamma$ subunits, as well as the $\alpha_3\beta_3\gamma$ assembly.
These  analyses  required  initial energy minimizations that distorted the crystal coordinates by
RMSD values of around 1\AA.
Cui and coworkers published the root mean square fluctations or RMSF per C$_\alpha$  of the  intra-monomer vibrations
 of the $\alpha$, $\beta$ and $\gamma$ subunits, as well as for the complex.
Each $\beta$ subunit, they concluded ``has the functionally relevant flexibility built into its structure". 
Our results reproduce the RMSF plots for the $\alpha$ and $\beta$ subunits, and indicate that
the FBM  patterns associated with the energy minimized 1BMF structures are largely shared by those of the current
PDB entry.  
Our analyses
add to the insights provided by the earlier studies by probing the nature and cause of the variable flexibility of the
$\alpha$ and $\beta$ subunits.  We continue efforts to develop a lexicon to describe nonlocal FBM where
the motility of any particular loop is causally linked with the motility of all parts of the peptide chain.
Highly mobile loops, it is seen, are not mobile in isolation: regions may be stiff locally and yet obtain high
temperature factors.
We examine the correlations of computed and observed temperature factors and note an interesting
signature suggesting the absence of one particular mode of vibration in an $\alpha$ subunit.

\subsection{Homunculus}

\begin{figure}[h]
\includegraphics[width=0.6\textwidth]{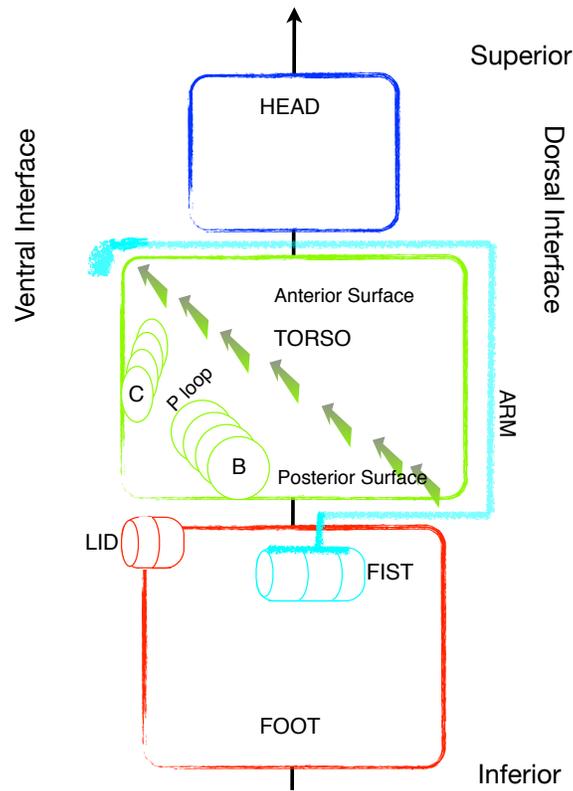}
\caption{Schematic representation of subunit. Each subunit consists of three domains, a head, torso and foot.
Viewed from the external surface, the axis of pseudosymmetry as indicated by the solid arrow, is behind the
subunit in this representation. Each subunit interacts with two neighbors in the hexamer: at the ventral surface
and at the dorsal surface.  The torso is divided into an anterior and a posterior surface region by a seven stranded parallel
$\beta$ sheet that extends from the inferior dorsal region to the superior ventral region. The posterior surface of the $\beta$
sheet consists of the nucleotide binding region including the P-loop and  helices B and C.  A shoulder region
at the interface of the head and torso near the ventral surface extends as an arm along the anterior surface
to connect to the foot domain as a short $\alpha$ helix, the fist.
}
\label{Figure02}
\end{figure}

To  characterize nonlocal FBM,  it will be helpful to adopt and extend a descriptive vocabulary. 
Figure \ref{Figure02} presents a schematic diagram of one subunit to provide a contextual lexicon for the regions  relevant to FBM.
The N terminal ``head'' domain, a six-stranded $\beta$ barrel extending from the N terminus to
residue $\beta$~Asp 77  and $\alpha$~Ile 94, is {\it superior} to the ``torso,'' the nucleotide-binding
domain consisting of a seven-stranded parallel $\beta$ sheet and associated $\alpha$ helices.
The C-terminal ``foot'' domain, a bundle of 6 ($\beta$ subunit) or 7 ($\alpha$ subunit) $\alpha$ helices is
{\it inferior} to the torso in this representation.  Each subunit forms two interfaces within the hexamer: 
with the central, rotor axis situated {\it internal} to the subunit,
a {\it ventral} surface involving the subunit's nucleotide-binding domain, as well as a {\it dorsal} surface
that abuts the neighboring subunit's nucleotide-binding region, are indicated in Figure 2.

The torso's central $\beta$ sheet extends diagonally from the inferior dorsal region to the superior ventral region.
The C-terminal, arrowed ends of the $\beta$ strands orient towards the internal surface  of the hexamer, at
a lower radial distance from the central axis than the N terminal ends of the $\beta$ strands which orient towards the
{\it external} surface of the hexamer.
The {\it anterior} surface of the torso's $\beta$ sheet faces the head region and obtains 4 $\alpha$ helices 
and associated loops.
The {\it posterior} surface of the  $\beta$ sheet  faces the foot and obtains two $\alpha$ helices (B and C) 
as well as the P-loop that binds nucleotide and cation.

The head domain links to the torso domain via a short ``neck''. 
A long linker ``arm'' extending from residues $\beta$~Asp 103 - Ile 137 and $\alpha$~Asp 116 - Ile 150 drapes
along the outer surface of the torso from the head region to the foot region where it connects to a
 firmly embedded short $\alpha$ helix, the ``fist,'' at residues $\beta$ 138-143 and $\alpha$ 151-155. 
 A ``shoulder'' region extending from the neck to the arm is situated superior to the torso at the ventral surface.
The superior, ventral region of the foot domain contributes either a highly mobile ``lid'' ($\beta$ Pro 417 - Pro 433) 
or a shorter ``strut'' ($\alpha$ Gln 430 - Glu 440) that serve to  either 
enhance or dampen oscillatory motions at the nucleotide-binding cleft.

\section{RESULTS}
\subsection{PDB coordinates}

We use the $\alpha$ and $\beta$ chains from coordinate file 2JDI \cite{walker07}. 
For this structure, mitochondrial F$_1$-ATPase from
bovine heart tissue was crystallized in the absence of preservatives and inhibitors and in the presence of ADP and
a non-hydrolyzable ATP analog AMP-PNP (abbreviated as ANP)  and  solved to 1.9\AA~resolution.
The 3 $\alpha$ subunits, chains A, B and C, as well as two $\beta$ chains, D and F,
each contain ANP and cation in their nucleotide-binding pockets while the remaining $\beta$ subunit, 
chain E, contains no nucleotide nor cation. 
Each chain folds to fit inside a wedge about 85\AA~along the central axis, 50\AA~in the radial direction,
and 55\AA~along the radial arc. 
By convention, each $\beta$ subunit is paired with the neighboring $\alpha$ subunit that abuts and contributes 
to that $\beta$ subunit's catalytic site: E with A, F with B and D with C.

In 2JDI the three $\alpha$ chains  adopt almost identical conformations. 
Their  all-atom superpositions
result in RMSD values of 0.6-0.7 \AA,  with  minor mismatches in
the alignment of either their head or foot  regions. The $\beta$ chains  D and F, both containing
ANP and Mg, obtain an RMSD of 0.6\AA.
Chain E, lacking nucleotide and cation,  differs from the other two $\beta$ chains by  3.8 \AA. 
The E chain's head and foot regions have swung to a higher radial distance from the central axis, creating
a more open or extended conformation. Superposing the individual head, torso and foot domains
of chains E and F  results in RMSD values of 0.2\AA, 1.3\AA~ and 0.5\AA,
indicating that the nearly 4\AA~ shift between these two chains is created to a large extent by  rigid body
rotations of these three domains.

The $\alpha$ chains extend from residues 24-510 and the $\beta$ chains from 9-474
(we maintain the PDB numbering convention that labels the N terminal $\beta$Ala residue as 1, not -4).
Chain C, as an example,  obtains 3715 atomic coordinates, whose 11145 internal, Cartesian degrees of freedom divide
among 3766 bond lengths, 4768 bond angles, 2605 dihedral angles and 6 rigid body degrees of freedom. 
We include only the ``soft'' dihedrals, 
including 469 $\phi$, 487 $\psi$ and 873 $\chi$ angles,  to study thermally induced equilibrium vibrations, 
reducing the available degrees of freedom from 11139 to  1829 while maintaining all bond lengths and angles to
PDB values.
In addition to the energies associated with these $dof$s,  chain C
obtains 22160 non bonded interactions (NBI) between all atom pairs
further apart than 3 bond lengths and less than a cutoff distance defined by the
inflection point of their van der Waal curves. The average NBI per atom is 5.3 and  includes roughly
the first shell of neighbors. The total  NBI divide between
9461 main chain-main chain interactions, 9222 main chain-side chain interactions, and 3477 side chain-side chain interactions.  
While we report on this distribution of dihedral angles and NBI, one might examine the effects of 
eliminating particular groups of NBI or soft dihedrals on the motility spectra, to assess, for example, the
effects of mutations on motility \cite{tirion15}. (In the isolated protein, the two different orientations of surface SC 
Arg 373  from chain A  do not affect the slow modes here described.)

\subsection{Normal Modes}

Normal modes were computed using ATMAN \cite{tirion14}. Thermal activations of modes were computed at 180K,
the temperature observed to divide harmonic from anharmonic motion in folded proteins \cite{petsko} and close
to the crystal temperature of 100K.  RMSF values per mode decrease rapidly with mode number, with
the first three modes contributing 64\% to the total, therefore our focus remains on these three softest, slowest 
modes \cite{tirion93}. 
Most analyses were 
carried out on chains C and F to study the distinct signatures of SUA and SUB.
Residue to residue comparisons between chains belonging to subsets $\alpha$ and $\beta$ used the Needleman \& Wunsch
sequence alignment algorithm available from the Protein Data Bank \cite{needleman}.
Results were checked for consistency using chains D and  A.
Chains B and E  both miss an 8 residue sequence in the foot ($\alpha$ 402-409, $\beta$ 388-395)  
in a region with  high experimental
temperature factors (over 60\AA$^2$  when the mean B-factor for chains A-F  is 16\AA$^2$).
To test for consistency with chains B and E, the missing 8 residues were built-in by rigid body
alignment of the missing region from a neighboring subunit.

\begin{figure}[h]
\includegraphics[angle=270,width=0.75\textwidth]{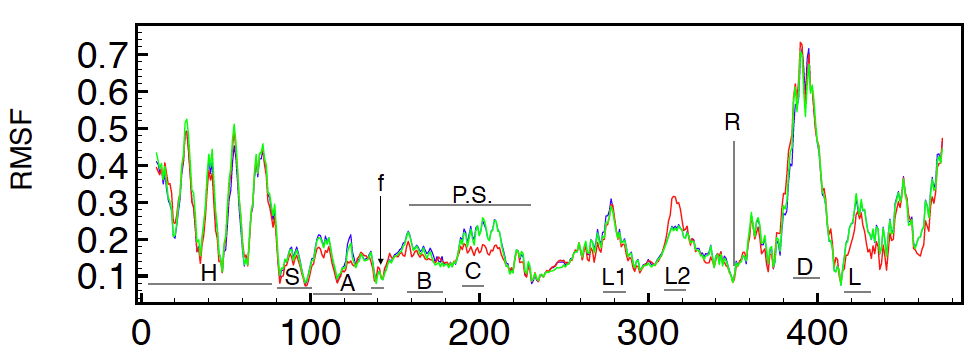}
\includegraphics[width=0.7\textwidth]{Figure03}
\caption{The computed RMSF in \AA~ per C$_\alpha$  due to the combined contributions of modes 1, 2 and 3 at 180K for
chains D (blue), E (red) and F (green).  
Horizontal bars identify regions: H head; S shoulder; A arm; f fist; B and C helix B and C; 
P.S. posterior surface of the torso's $\beta$
sheet; L1 loop $\beta$Pro 276-Pro 283; L2 loop $\beta$Pro 313-Pro 322;
R identifies location of the Arginine finger ($\beta$356) that separates the torso and foot domains; D residues 390-400  include
the mobile DELSEED region; L the mobile lid ($\beta$ Pro 417-Pro 433)
that forms  the posterior surface of the nucleotide binding cleft. Interestingly, the
regions close to and likely to interact with the $\gamma$ rotor (D, L1 and L2) obtain relatively high mobility, even
in absence of the rotor.
}
\label{Figure03}
\end{figure}

In all, the computed modes describe the same motions, both among chains A, B and C as well as among chains D, E and F.
For example, Figure \ref{Figure03} 
shows the RMSF per C$_\alpha$ due to the combined effects of modes 1, 2 and 3 for chains D, E and F, the
three $\beta$ subunits.
The computed motility profiles for these 3 chains superpose almost perfectly, with chain E presenting with slightly
different RMSF amplitudes in the region that includes  helix C (residues $\beta$190 - 215) 
as well as a loop on the anterior side of the torso's $\beta$ sheet (residues $\beta$317-321).
These plots closely match the equivalent RMSF per C$_\alpha$ plot for the unliganded $\beta$ subunit, Figure 10a, of Cui and coworkers \cite{cui}, with the mismatch in relative amplitudes due to our use of 180K for the activation temperature and 
our use of a small subset of all modes.

 While the similarity of the slow modes amongst the $\alpha$
chains is not  surprising due to the close structural similarity between these chains, 
the match of the slow modes amongst the different $\beta$ chains is reassuring. The E chain
adopts a conformation distinct from chains D and F, and the similarity of the three $\beta$ chain mobility profiles 
provides a strong indication
of the intrinsic character of these innate $dof$s. 
One might therefore expect  eigenmodes to contribute to the
interconversion of these different conformations. 

As discussed in the Introduction,
the current analyses pertain primarily to the computed eigenspectra of the $\alpha$ and $\beta$ subunits 
in absence of bound ligands.
The $\beta$ subunits transition between open and closed conformations as reactants enter
and products leave the catalytic site.  Furthermore, 2JDI reports 351 $\pm$ 16 crystal water
molecules for the three $\alpha$ subunits,
and 351 $\pm$ 85 crystal waters for the three $\beta$ subunits, indicating a large variation in the numbers of crystal waters 
accompanying the structural transitions. 
To assess the extent to which ligands affect the expression of the slowest modes, we computed the
eigenspectra and eigenvectors of the protein plus nucleotide and cation; protein plus tight bound waters; and protein plus
nucleotide, cation and tight bound waters. A water molecule was considered tightly bound if it obtained atleast
5 NBI with protein atoms.
We find that the presence
of nucleotide and cation slightly shifts the frequencies of the slowest modes, making them  stiffer and resulting
in slightly smaller amplitudes of oscilllation at any given temperature.  The presence of tightly bound waters
slightly alters the slow eigenvector shapes, tending to enhance the propensity seen
in the $\beta$ subunits to open and close the catalytic site. 
As the effects of ligands on the slow modes as modeled by L79  
do not result in significant differences, we here focus our research on the isolated protein chains.

\begin{figure}[h]
\includegraphics[width=0.7\textwidth]{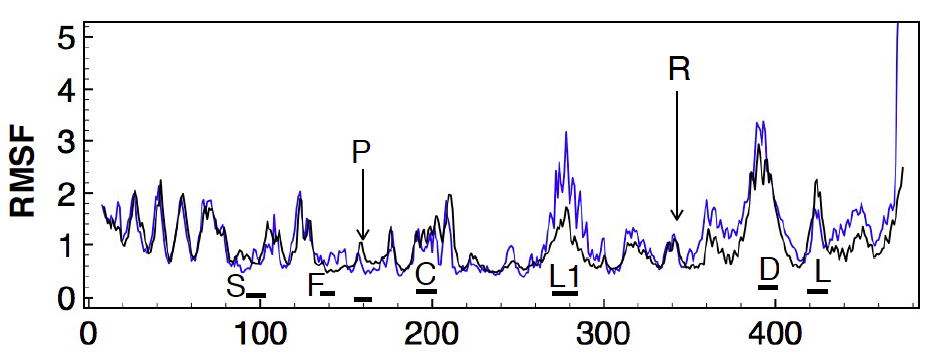}
\caption{The RMSF in \AA~ per C$_\alpha$ obtained from a 100ns MD simulation of hydrated, 
isolated and liganded subunit 
$\beta$ (chain F) in blue \cite{enrique} 
and of the same chain, unliganded, computed from the slowest 50 normal modes in black.
The close overlap indicates
that the long term motility of the polypeptide chain is well captured by the harmonic analysis, with only slight mismatches
in the magnitude of peaks associated with the fist region and P-loop region, likely due to the absence of nucleotide
in the PDB-NMA; a different magnitude for the L1 peak near the $\gamma$ rotor axis; and  some
mismatch in magnitude in the region immediately after the arginine finger and after the mobile lid region. 
Both the DELSEED loop as well as the mobile lid region in the foot domain overlap closely.
}
\label{Figure04}
\end{figure}

The use of unliganded protein chains is supported by  earlier analyses
of the liganded and unliganded motility profiles of $\beta$ subunits \cite{enrique,cui,grubmuller} which show
RMSF per C$_\alpha$ shifts only in the magnitudes of certain peaks. Such shifts imply changes in the amplitudes of motions,
not changes in the character of the motion. For example, Hahn-Herrera and coworkers recently published results
of MD simulations of isolated $\alpha$ and $\beta$ chains of PDB entry 2JDI \cite{enrique}.  In addition to the protonated PDB
coordinates, ATP and cation, his group included 40,338 water molecules in the system. After 10 ns of
equilibration, the system  underwent 100 ns of unbiased MD simulation.
The resultant RMSF values of the C$_\alpha$ atoms of chain F (liganded $\beta$ subunit) 
is shown as the orange line in Figure \ref{Figure04} (data kindly provided
by Prof. Garcia-Hernandez). Current NMA  predictions due to the slowest 50 modes for the same chain, unliganded and
up-scaled by a factor of 2.6, are superposed with the black line. Harmonic oscillations at 300K rather than the 180K
used here, increase the amplitudes of oscillation by a factor of $\sqrt{300/180}$ or 1.3.  Furthermore, anharmonic contributions
in MD are thought to double predicted displacements compared to harmonic NMA \cite{kitao} and may explain
this scaling factor.
The motility patterns for the head, shoulder,
arm and posterior surface of the torso domain match closely. The peaks associated with those regions closest to
the central $\gamma$ axis, such as the DELSEED loop, are also matched, though with slight mismatches in their magnitudes.
By and large, therefore, 
the long-term, 100 ns,  motility profile of  a solvated protein chain
derived from a detailed and accurate force field matches that predicted by PDB-NMA. 
This match provides a strong indication that long term motility is the result of correlated motions and not
the result of separate regions moving independently. 
The NMA profile,
produced by the superposition of all the modes of oscillation,  may be separated into and studied as  individual modes
in an effort to characterize and apprehend the source of these long-term motility profiles.

\subsection{The first three modes}

\begin{figure}[h]
\includegraphics[width=0.8\textwidth]{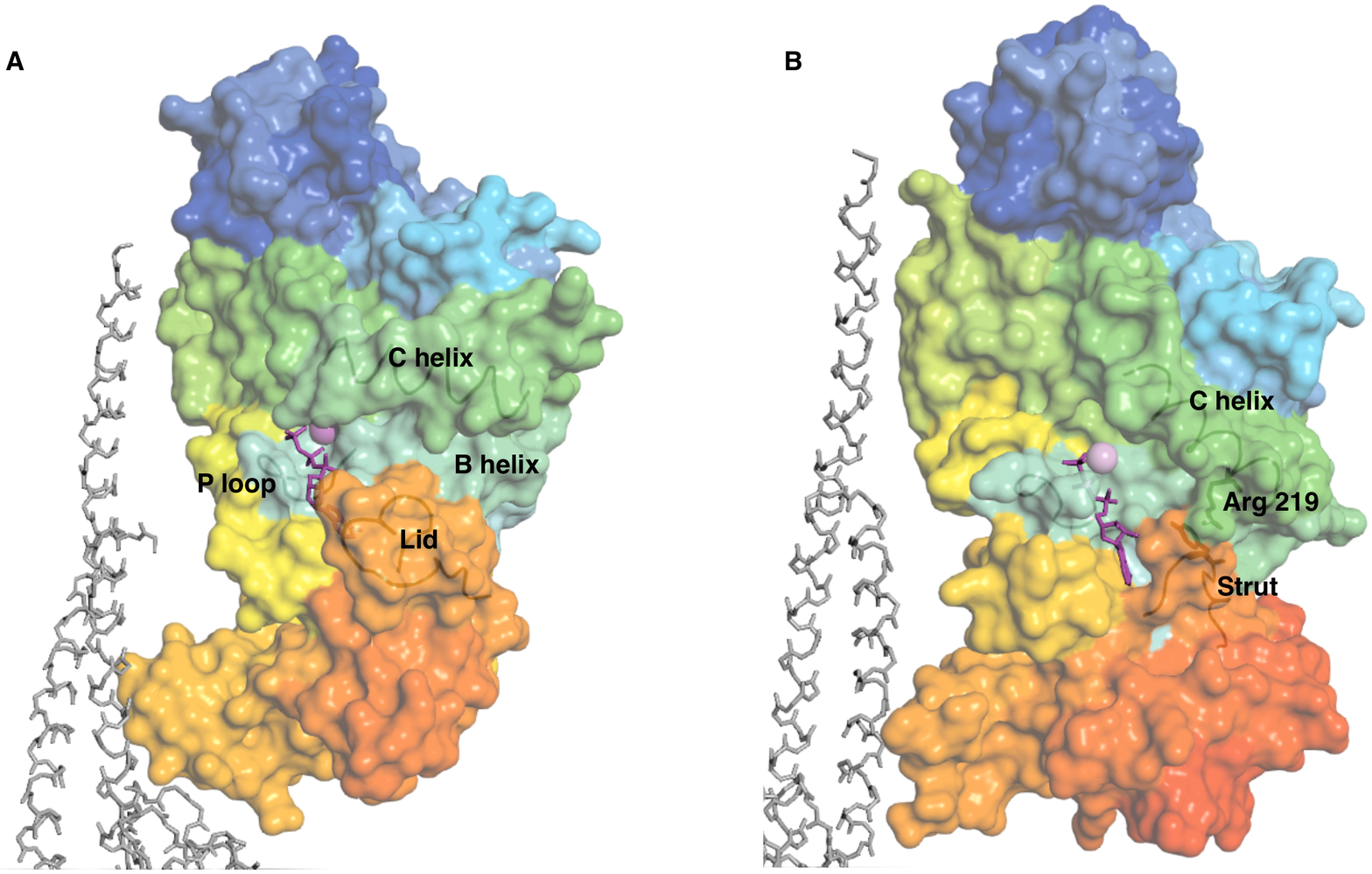}
\caption{Surface representations of the ventral surfaces of SUB (left) and SUA (right) showing the
nucleotide binding clefts with ATP colored magenta and Mg as a pink sphere.
The N and C terminal domains of subunit $\gamma$'s  are shown in a gray stick representation.
SUA and SUB are colored with a rainbow scheme with the
N terminal, head domain in dark blue, the shoulder and fist (visible beneath the strut in B) in cyan 
and the C terminal foot domains orange and red.   The P-loop, to the left of ATP, extends as helix B
in light blue and forms the base of the nucleotide binding cleft. The top of this cleft is formed by helix C and
the bottom of the cleft by the mobile lid (A) or the rigid strut (B). In SUB  helix C and the  lid 
experience pronounced swinging motions in eigenmodes 1, 2 and 3,  motions  absent in SUA where strut residue
Tyr 433 forms extensive steric interactions with helix C residue Arg 219. Figures prepared in PyMol.
}
\label{Figure05}
\end{figure}

The slowest three modes of SUA and SUB  obtain similar profiles for the
FBMs involving relative displacements of the head, torso and foot regions.
While computed using dihedral angles, the slowest three modes are also perpendicular in Cartesian
space, with oscillations of mode 1 about an axis aligned along a radial direction; of mode 2 about a radial arc and  
for mode 3 about a direction parallel to the central, $\gamma$ axis. 
While the oscillations described by each mode appear largely similar between SUA and SUB, 
their effects on the posterior side of the $\beta$ sheet, and in particular on the cleft that ends
at the Walker A motif or P-loop (GxxxxGKT, residues $\alpha$ 169-177 and $\beta$ 156-163)
that coordinates the $\beta$ phosphate of the nucleotide, is very distinct.
In SUB this cleft extends  approximately 20 \AA~ along a radial direction to the external surface with
its base  formed by  helix B extending from the P-loop (Figure \ref{Figure05}A). 
The top of the cleft is formed by  helix C and the bottom
by the lid, with $\beta$Phe 424 aligned with the ribose ring. This  cleft in SUB 
displays pronounced opening and closing motions for modes 1 and 2 and
a relative, grinding motion in mode 3, motions completely absent in the $\alpha$ subunits.
The mobile lid is not present in SUA
where the strut blocks
relative motion across the crevasse, with foot residue
$\alpha$Tyr 433 forming extensive steric interactions with $\alpha$ Arg 219 from the top of the cleft (Figure \ref{Figure05}B).  

\begin{figure}[h]
\includegraphics[width=0.7\textwidth]{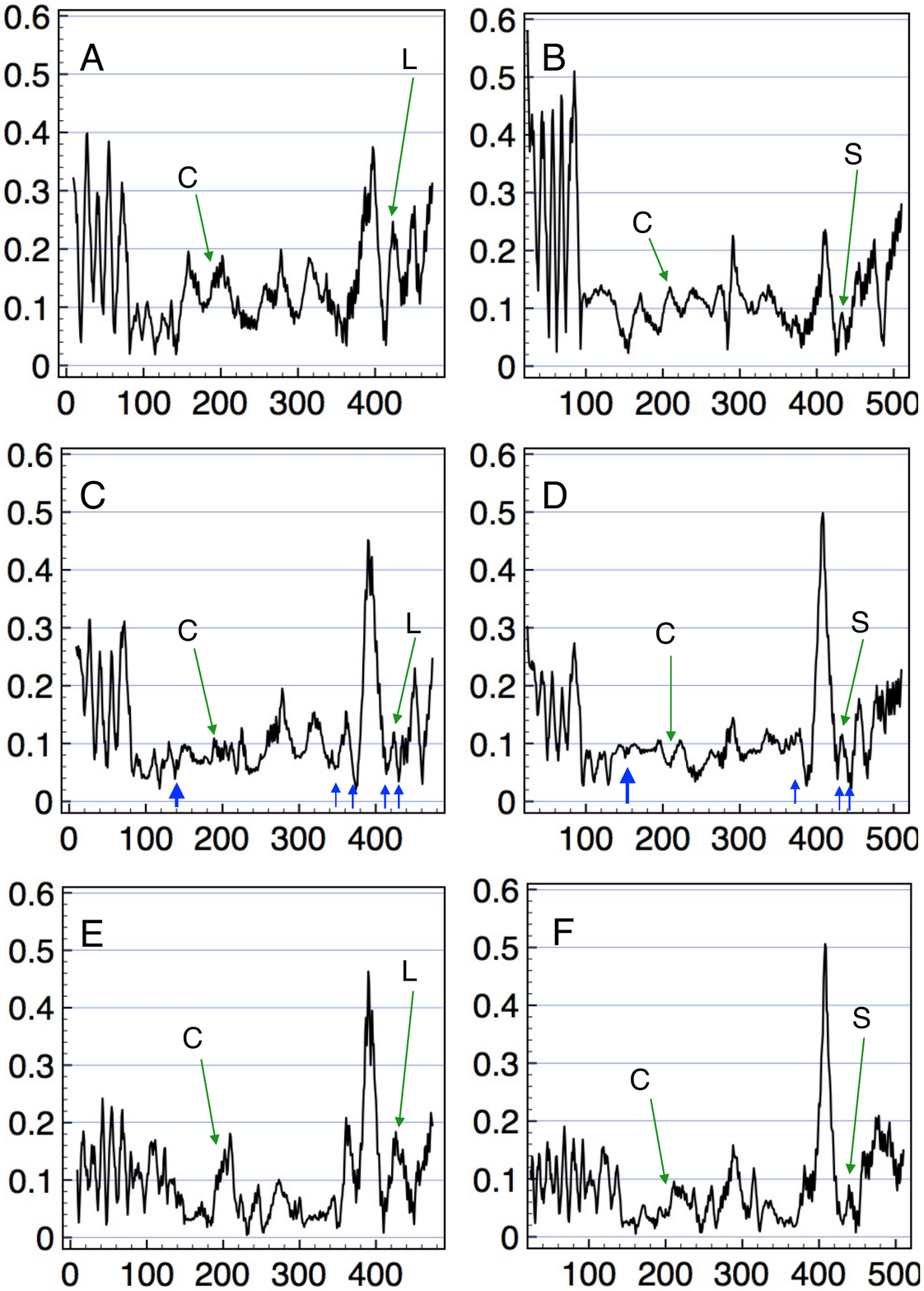}
\caption{RMSF in \AA~ per C$_\alpha$ for mode 1 (A and B), mode 2 (C and D) and mode 3 (E and F) with left images 
for SUB and right images for SUA.  Peaks associated with helix C and either the lid  L (SUB) or the strut S (SUA) are
indicated by slanted green arrows. The short vertical blue arrows indicate regions of low motility near to the fist residues.
 and demonstrate greater motility in SUB than SUA.
Mode 1, a sidewise rolling of the head and foot domains about an 
axis roughly aligned with
helix B, exhibits  reduced head motility and greater foot motility in SUB
compared to SUA. The increased motility at the posterior surface of the torso's $\beta$ sheet
 appears as a chewing type motion across the nucleotide binding cleft. 
Mode 2 is a flexing towards and away from the central axis of the head and foot domains about an axis aligned with the fist helix
with the torso relatively immobile. That portion of the nucleotide binding cleft furthest removed from the fist-helix, namely
the N terminal end of helix C, experiences a small  spike in SUB not present in SUA and again pertains to the greater binding cleft
motility seen in this subunit.
Mode 3 presents as a twist type motion of the head and foot
domains where, like mode 1, the helix C peak  and the lid/strut peak are more pronounced in SUB than SUA, and
present in SUB as a grinding type motion across the binding cleft.
}
\label{Figure06}
\end{figure}

The slowest modes of SUA and SUB (Figures \ref{Figure06}A and \ref{Figure06}B),
with frequencies of $\alpha$ 2.6 cm$^{-1}$ and $\beta$ 2.7 cm$^{-1}$,  pertain to a sidewise rolling,
with the head and foot regions rocking towards each other along a radial axis closely aligned with the P-loop helix (B). 
The arm in SUB
seems to function as a caliper, with the $C_\alpha$s of shoulder residues Asp 103 and fist Gly 136
separating  33.2 $\pm$ 1.2\AA. This results, on the other side of the torso's $\beta$ sheet, in a sizable fluctuation
of nearly 3\AA~between the top of the cleft and the mobile lid, in an
up-and-down chewing type motion.
In SUA, the rolling of the head towards and
away from the foot along a radial axis  is again observed.
However, the arm in SUA does not function as a caliper, and the
equivalent $C_\alpha$ atoms at residues Asp 116 and Gly 149 remain at a nearly steady 28.5$\pm$0.2\AA~separation during
thermal activation.
As a result, the cleft and access to the cleft leading to the P-loop is not distorted. 

Modes 2 (Figures \ref{Figure06}C and \ref{Figure06}D), both
with frequency of  3.1 cm$^{-1}$, pertain to a flexing toward the central axis, 
with a short stationary  $\alpha$ helix oriented along a radial
arc marking the axis about which the top and bottom portions of the chain oscillate. This short helix 
 is the fist that interconnects the head and foot regions,  moving in tight synchrony
with the foot domain due to extensive packing interactions, including  SUB residues
Lys 138, Val 139 and Leu 142 and Leu 143. 
 The NZ of Lys 138, for example, maintains fixed distances of separation
 with the main chain carbonyl oxygens of  Arg 142, Phe 457, Gly 461, Phe 413 and Val 460 during thermal activation.
The effect, as seen in Figure \ref{Figure06}C, is startling: the fist residues $\beta$ 138-143 (indicated by the short, left-most
vertical arrow), obtain low RMSF values along with 
those nearby regions of the foot indicated by the remaining vertical arrows. 
Several of the $\alpha$ helices in the foot radiate away
from the fist region, with those regions of the helices near the fist nearly immobile while the  remainder of these helices
obtain significant RMSF values. For example, the C terminal region of helix  $\beta$399-414 and the N
terminal region of helix $\beta$434-448 obtain small RMSF values (right most arrows), while their connecting
loop forms the mobile lid motif.
As a result, the head and foot regions
simultaneously rock towards and away from the central $\gamma$ axis, with the nucleotide binding cleft
experiencing a more pronounced opening nearer the P-loop rather than near the outer edge as in mode 1.
The second mode, as will be seen,
contributes the largest amplitude when the slow modes are used as $dof$s to transform the coordinates
of the D and F chain to the E chain. 

Modes 3  (Figures \ref{Figure06}E and \ref{Figure06}F),
with computed frequencies of $\alpha$ 3.8 cm$^{-1}$ and $\beta$ 3.6 cm$^{-1}$, pertain to a 
twisting motion of the head and foot regions about an axis roughly parallel
to the central $\gamma$ axis.  
This motion reveals considerable swinging of that portion of the foot that includes the DELSEED region
nearest the central $\gamma$ axis, and is coupled to an appreciable relative twisting of the top
and bottom of the nucleotide binding cleft in SUB. As an example, the 11.2\AA~  distance of separation in SUB 
between lid atom Phe 424 CZ  and
Arg 189 CZ at the N terminal end of helix C,  reduces to 10.4\AA~before  increasing to 12.3\AA~
during one cycle of oscillation, while the 16.5\AA~distance of separation between the same Phe 424 CZ  and  Glu 202 CD at the
C terminal end of helix C, first increases to 18.0\AA~before  reducing to 15.0\AA~during the same
cycle of oscillation.
This motion seems enabled by the torso's central $\beta$ sheet as the twist angle between 
adjacent strands of the $\beta$ sheet varies slightly as the head and foot domains rotate in opposite sense \cite{orozco}.
 As before, this distortion of the nucleotide binding cleft is not seen in SUA.

We next examine the correlations of the computed motilities 
against the  experimentally determined temperature factors before computing to what extent these three
orthogonal modes reduce the RMSD between the closed and open $\beta$ monomers. In the discussion, we will consider
the sources of the variable stiffness characteristics in SUA and SUB.

\subsection{Correlations with crystallographic temperature factors}

\begin{figure}[h]
\includegraphics[width=0.7\textwidth]{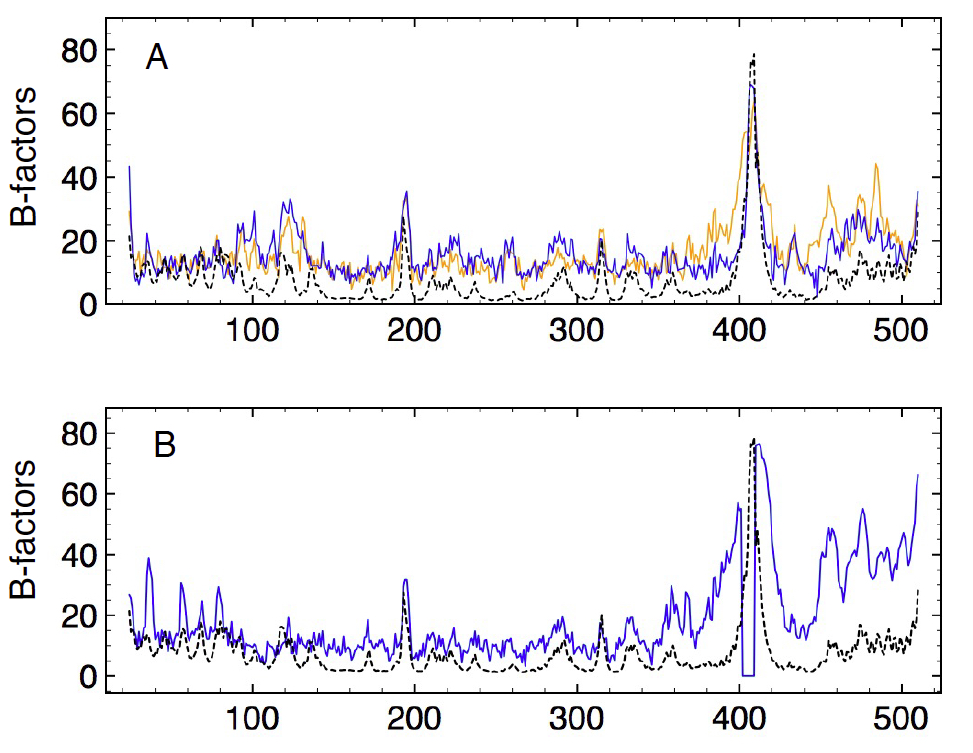}
\caption{Crystallographic B-factor in \AA$^2$ per C$_\alpha$ for the three  $\alpha$ subunits: chains A and C in orange
and blue in top panel and chain B in blue, bottom panel. The PDB entry for chain B misses residues 402-409.
Superposed in broken black line, the unscaled computed
B-factors derived from the sum of modes 2-50, top panel, or 1-50, bottom panel. The  experimental B-factor
plot for chain B with the distinct head peaks is better reproduced by excluding mode 1 in the sum.
}
\label{Figure07}
\end{figure}

We were interested to see how well the crystallographically determined  Debye-Waller  or B-factors 
for the three $\alpha$ chains as well as  for the three $\beta$
chains compared, in addition to their similarities to the computed motilities.  The 3 $\alpha$ chains obtain
all-atom RMSDs of around 0.7\AA, and one might expect, barring crystal packing effects, very similar
B  plots. In fact, the crystallographic   B-factors per C$_\alpha$ for chains A and C are
quite similar (Figure \ref{Figure07}A, solid lines), as expected, 
but chain B obtains a distinct signature (Figure \ref{Figure07}B, solid line), with 5 pronounced peaks in the head region,
a reduced amplitude in the neck and shoulder region, and enhanced motility in the foot region. 
Superposed on these plots with dotted lines are the {\it unscaled} 
computed temperature factors of the first 50 modes (Figure \ref{Figure07}B, dashed line)
and of modes 2-50 (Figure \ref{Figure07}A, dashed line).  The experimental temperature signature of chain B is  reproduced
reasonably well by the contributions of the slowest 50 modes, with its distinct head peaks, helix  C and
DELSEED peaks as well as several other other torso and foot peaks accounted for.
Intriguingly, the experimental  temperature factors
of chains A and C are best reproduced by excluding mode 1 in the sum, which similarly eliminates the distinct peaks
in the head region.
One interpretation of these data is that in the 2JDI crystal, 
chain B vibrates along all eigenmodes while chains A and C
lack the  slowest degree of freedom, perhaps due to their differing interaction with the central rotor proteins.
The current computations consider only intra-monomer, not inter-monomer, packing
interactions,  and the possibility  that the source of the variable experimental Debye Waller signatures
for the three $\alpha$ chains might arise from
variable symmetry axes activation is a novel concept  for molecules as large as proteins.
Laser spectroscopists have successfully activated single resonances (eigenfrequencies) in 
multi-atom systems in order to alter reaction rates \cite{hoffmann,zare}; perhaps 
 the altered B chain Debye Waller factor profile
in the crystal structure map is indicative of selective eigenmode damping.

\begin{figure}[h]
\includegraphics[width=0.7\textwidth]{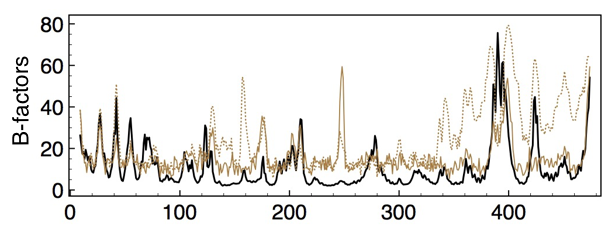}
\caption{Crystallographic B-factors in \AA$^2$ per C$_\alpha$ for the   $\beta$ subunits: chains D and F obtain
similar B-factor profiles shown by solid brown curve and chain E with dashed brown curve.  The unscaled
theoretical values are superposed in solid black. Unlike the D and F chains whose experimental B-factors
are reasonably well modeled by PDB-NMA, the distinct experimental B-factor curve of chain E is less
well modeled.  Each crystalline hexamer unit obtains one unliganded, $\beta$ subunit with altered solvation
characteristics due to its open configuration. Reduced statistics combined with possible greater variability
in conformation of foot domain, including the fist, might explain the mismatch.  
The peak at 245 corresponds to a loop on the
anterior surface of the torso's $\beta$ sheet, at high radius, and is  involved perhaps with inter-hexamer packing
interactions.  The enhanced P-loop peak (156-163) motility is likely due to the absence of nucleotide
and cation.
}
\label{Figure08}
\end{figure}

The experimental temperature factors for chains D, E and F are shown in Figure \ref{Figure08}. The two structurally similar
subunits, D and F, display similar vibrational patterns indicated by the solid brown curve,
while chain E obtains significantly higher B-factors for the C-terminal residues after the torso's final $\beta$ strand, 
as well as an additional spike for those residues connecting the fist to the P-loop, as indicated by the dashed curve.
 Superposed on these curves in black are
the  computed B-factors due to the slowest 50 modes of chain F.
The theoretical values reproduce some of the experimental features, obtaining similar head and foot motility
profiles yet notably missing   several peaks in the torso region. For example,
 the experimental peak at residues 238-242, a loop on the anterior surface of the torso's $\beta$ sheet, is not predicted.
As discussed, the computed motility profiles for chains D, E and F are almost indistinguishable, hence the
source of the E chain's variable B signature is likely due to effects other than intra-minimum, equilibrium vibrations.


While experimental temperature factors include many effects other than the intra-monomer vibrations
considered here, such as harmonic inter-monomer and crystalline vibrations as well as anharmonic noise, 
nonetheless it might be
instructive to determine whether these data support the central assertion of reduced motility at the posterior surface
of the torso's $\beta$ sheet in SUA compared to SUB.  PDB-NMA predicts the C$_\alpha$ atoms in SUA and SUB to obtain
 average B-factors of 9.7$\AA^2$~ and 10.9$\AA^2$.  The C$_\alpha$ atoms  posterior to the $\beta$ sheet,
$\alpha$169-229 and $\beta$157-214, obtain average theoretical B-factors of 6.2$\AA^2$~ and 9.9$\AA^2$, indicating
the reduced mobility of this region in SUA compared to SUB.  The average experimental C$_\alpha$ B-factors
for SUA and SUB are 15.8$\AA^2$~ and 15.9$\AA^2$.  The average experimental B-factors for the C$_\alpha$ atoms posterior to
the $\beta$ sheet are 15.5$\AA^2$ and 18.8$\AA^2$, indicating once again that the posterior surface of the torso's $\beta$ sheet
obtains lower motility character in SUA than in SUB. 
In sum, the experimental B-factors do not contradict
the predictions of NMA;  support the observation that the nucleotide binding region experiences reduced motility; 
 suggests the B-factor data of the  $\alpha$ subunits to be more accurately modeled
by NMA than those of the $\beta$ subunits; and may indicate selective modal damping in an $\alpha$ subunit.

\subsection{Interconversion of SUB structures with modes}

\begin{figure}[h]
\includegraphics[width=0.7\textwidth]{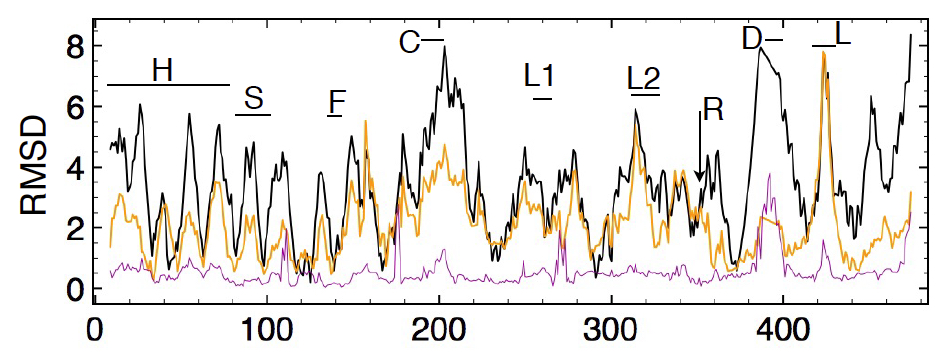}
\caption{RMSD in \AA~ per C$_\alpha$ between the PDB coordinates of chain E (open conformation) and
chain F (closed conformation) in black, with an overall RMSD of 3.8\AA.   Horizontal bars identify regions
as defined in Figure \ref{Figure03}.
A similar curve for the RMSD between
chain E and the normal mode deformed chain F, shown in orange, obtains an overall RMSD of 2.3\AA.
For comparison, the RMSD curve for the similarly shaped closed chains D and F, with an overall RMSD of 0.6\AA,
is shown in purple.  These data show
that the initial large mismatch in the alignment of the head, shoulder and arm residues as well as the foot residues
after the arginine finger, improves significantly after adjusting chain F coordinates along the slowest three modes.
The fit of the lid (L) region does not improve, suggesting a nonharmonic contribution to the shift of this
region during the interconversion of the open and closed conformations.
For the torso, there is improvement in the fit  in that region on the posterior surface that includes the helix C residues. 
The fit of the remainder of the torso region is not much improved, indicating that shifts within this region are also less well 
modeled by harmonic, intra-minimum oscillations exclusively. PDB 2JDI chain E lacks residues 388-395 within the
DELSEED region.
}
\label{Figure09}
\end{figure}

By deforming the coordinates of SUB chain F along the directions of its three slowest modes, we tested to what
extent those  coordinates could reduce the RMSD between chains F and E (Figure \ref{Figure09}). 
The initial RMSD of all mainchain heavy atoms of 3.8\AA~is reduced to 2.3\AA~using relative contributions
of 25\%, 45\% and 30\%~ of modes 1,2 and 3. 
The resultant structure shows that the
transformation aligns the relative positions of the head, foot and torso domains but does not much improve the
alignments within domains, that remain at 0.25\AA, 1.2\AA~and 0.53\AA~RMSD from those of chain E.
For example, alignment of the crystal F and E chains shows a distance of separation of their C termini of
8.5\AA~ that reduces to 1.5\AA~ for the normal-mode-deformed F chain. Likewise, the N termini in the crystal
coordinates are 4.5\AA~apart, which reduces to 2.4\AA~ for the NM-deformed chain. Within the torso domain, however,
only slight improvements to the RMSD are achieved, with observed realignment of the nucleotide binding cleft
not modeled as well by the slowest three modes. The cleft separation, for example,  from C helix C$_\alpha$ 203 to mobile lid
C$_\alpha$ 420 in chain F is 13.8\AA~ and in chain E is 5.1\AA.  The NM-deformed F chain reduces this to
 11.1\AA. 
 
A similar analysis of chain D reduces an initial RMSD to chain E of 3.8\AA~ to 3.6\AA~with use of 8.5 kT of mode 1,
to 2.9\AA~ with 17.3 kT of mode 2, and to 2.7\AA~with 8.5 kT of mode 3.  
The importance of natural or innate flexibility in effecting the open to closed transformation of $\beta$ subunits
was noted by B{\"o}ckmann and Grubm{\"u}ller, based on MD simulations of unliganded, open $\beta$ subunits. They
found fast, spontaneous and nucleotide-independent closure of the open $\beta$ subunit, with changes not
localized to the nucleotide binding region and not exerted from adjacent $\alpha$ subunits: ``the main driving
force for the closure is internal to the $\beta$-subunit" they concluded \cite{grubmuller}.


\section{Discussion}

\subsection{Design and Flexibility of Subunits}

SUA and SUB present very similar topologies  and folds  leading to  similar flexibility profiles.
A long linker arm straddles a central torso domain to interconnect similar head and foot domains.
A seven stranded parallel $\beta$ sheet spans the torso from the ventral surface near the head to
the dorsal surface near the foot, and divides the torso into an anterior region involved with
inter-subunit packing  and a posterior region that includes  the nucleotide binding cleft. Both have
an important short $\alpha$ helix at the end of the linker arm embedded tightly in the foot domain.
This helix  forms
 nonbonded interactions with  residues forming the torso's eighth, antiparallel $\beta$ strand
(in SUB the pattern of inter-strand hydrogen bonding precludes its designation as a $\beta$ strand);
 the N terminal end of the  long $\alpha$ helix after the arginine finger; the C terminal end of the foot's
third $\alpha$ helix and the N terminal end of its fourth $\alpha$ helix whose connecting loop constitutes
the mobile lid or the strut; as well as with the C terminal end of the torso's  helix B. 
The fist, in other words, while still near the N terminal end of the polypeptide chain (residues 138-143), also
coordinates the motion of foot residues, including the critical lid or strut, as well as  torso residues. 
Specifically, as indicated in
Figure \ref{Figure06}C and D, these regions, like the fist residues, obtain very small motilities, 
indicating the critical importance of the fist in coordinating the motions while maintaining
the structural integrity of the chain's fold.

This design leads to similar motility profiles, with the softest motilities associated with
a side-wise rolling, a flexing towards the central axis, and a twisting about the central axis of the
head and foot domains.  While the inter-domain motions of the head, foot and torso appear
largely similar between SUA and SUB, important differences exist between the motility profiles
of the region posterior to the torso's $\beta$ sheet: the nucleotide binding region. In SUB the
nucleotide binding cleft exhibits pronounced opening and closing motions, motilities
absent in SUA, and points to an unsurprising difference between enzymes and structural proteins.
Enzymes require entry and exit of reactants and products in a manner that predicates reaction
rates, whereas structural proteins have no need of such intra-domain motilities.  
What design features allow one chain to be an active enzyme, the other not?

\subsection{Source of Variable  Torso Stiffness}

In SUA the residues forming the foot domain move in lockstep with the fist residues and also with all
the residues comprising the posterior surface of the torso's $\beta$ sheet. In SUB the residues forming
the foot and fist move together with only  helix B of the posterior surface:  helix C moves {\it en masse}
with the shoulder and residues of the anterior surface. Why?
Certainly the presence in SUA of the foot's strut element and the resultant tight packing
between strut residue $\alpha$Tyr 433 and helix C residue $\alpha$Arg 219 creates a significant obstruction to intra-cleft motilities. 
But in addition to this localized distinction,  distributed differences exist between SUA and SUB in terms of the distribution
of NBI.  The 35 arm residues ($\alpha$ Asp 116 - Ile 150 and $\beta$ Asp 103 - 137) for example, pack more tightly
in SUA, with 1186 NBI between  arm  atoms and non-arm atoms, and 789 such NBI in SUB. Also, as pointed out in the
Results section, the diagonal distance across the torso as measured by the distance of separation of the C$_\alpha$ atoms
of the first and last arm residues is 33\AA~in SUB and 28\AA~ in SUA, another indication of a more compact torso 
arrangement in SUA.
The less tightly packed arm segment of SUB obtains two additional elements of secondary structure:
a 310 helix (Phe 123 - Glu 125), as well as a short antiparallel $\beta$
sheet preceding the fist (Ile 132 - Leu 133 and Tyr 146 - Ala 147), both absent in SUA.

Another element that seems critical to the distinct motility profiles centers on the interface of
the foot and torso domains and pertains to the fold of the chain immediately after $\alpha$ Arg 373 and
$\beta$ Arg 356.  
This so-called arginine finger is equivalently situated in SUA and SUB: contributed by the foot domain
but oriented towards the anterior surface of the $\beta$ sheet, at the dorsal surface between the torso and the arm.
In SUA the chain after Arg 373  turns away from the dorsal interface  and inserts
into the cleft formed at the junction of the arm and fist where the two $\beta$ strands preceding the fist meet. Likely
this is the reason that SUA cannot form a $\beta$ structure here:  any potential $\beta$ strand  contributed by the arm segment
is displaced  by this loop after the arginine finger. 
This tight packing in SUA effectively creates a block structure where fist and foot motility transfers {\it en masse} to
helix B and, due to the strut, to helix C.
 In SUB the residues after Arg 356, instead of folding away from the dorsal surface, fold towards it,
creating the space that allows the formation of the locally stiff antiparallel $\beta$ strand that precedes the fist. 
This arrangement eliminates the tight packed, block-like structure and creates the possibility of intra-domain motion,
with helix B moving with the foot domain, and helix C moving as part of the shoulder and anterior surface of the torso.

The eigenmodes of SUA and SUB, therefore, demonstrate  the two qualities that play off each other
so that two chains with identical folds have different motilities: spaciousness  provided
by regions with relatively low density of NBI against local stiffness elements, such aditional elements
of secondary structure. 
Lack of tight packing in SUB is supplemented not only by the presence of additional  rigid elements of secondary
structure, but also by inclusion of additional prolines. SUB has 23 while SUA has
17 of them.   
Rigid body motion seems enabled by distributed, high density of NBI forcing such 
regions to move {\it en masse}.
Relative motion within a domain or region requires sufficient space to enable  the various
elements to move in an opposing sense. Such spaciousness likely could result in localized fraying or unfolding of the polypeptide
chain, were it not offset by these elements of local stiffness.
In both cases there is effective and fine-tuned transmission of stiffness throughout the chain.
The distributed nature of the ``control'' of mobility characteristics minimizes likelihood of disruption but also permits
higher precision. For example, 
the absence of the strut element in SUB permits motility within
the nucleotide binding cleft, but this additional degree of freedom is not disorderly or haphazard: many
nonlocalized, distributed elements conspire to control the expression of this degree of freedom.

\subsection{Intra-monomer oscillations within assembly}

Eigenmodes were computed for isolated $\alpha$ and $\beta$ chains without regard to the additional 
inter-monomer NBIs in the assembly, hence steric clashes were expected when these intra-monomer modes were 
activated within the assembly.  Unexpectedly, steric inter-monomer clashes are avoided in the assembly
when the isolated-monomer modes were activated simultaneously. 
The interfaces maintain their interdigitation
during activation of each (intra-monomer) mode, with the projections of one subunit
maintaining closely similar dispositions with the concavities of its neighbor. This effect seems to
be caused by the relative
immobility of those residues involved with inter-monomer interactions.

In particular, a ring of  connectivity that extends around the assembly at the level of the superior surface of the torso  
 maintains  close interdigitations, with shoulder elements of one subunit packing with arm elements of the neighboring subunit.
Specifically, at the dorsal surface of SUA, arm residues $\alpha$ Pro 134 - Pro 138 form a loop construct that 
latch tightly with the neighbor's shoulder loop $\beta$ Pro 101 - Pro 107, with $\alpha$ Ile 136 - Ile 137 buttressed
between the neighbor's shoulder  loop and the N terminal (inner) end of its  helix C. 
Both interdigitating loops at this dorsal surface are braced by proline residues that provide the requisite rigidity 
to these latch elements.

At the ventral surface of SUA, meanwhile, shoulder loop residues $\alpha$ Ala 114 - Pro 120 form the loop
construct that together with helix C latch onto the neighbor's $\beta$ Pro 121 - Glu 125 arm residues,
with the protrusion of $\beta$ Phe 123 maintaining a tightly coordinated orientation with the neighboring subunit
during thermal activation.
  Note that one of the two bracing prolines in the SUB's arm loop is replaced
 by a short 310 helix at residues $\beta$ Phe 123 - Glu 125 that seems to confer the requisite rigidity 
to help maintain hexamer integrity.

Activating the oscillatory motion of each mode simultaneously therefore suggests a 
relatively immobile anterior torso surface coordinating
hexamer stability while the foot and posterior torso surface  seem ``free'' to
oscillate independently of its neighbors.
This feature, of isolated, individual components comprising an assembly already possessing the flexibility 
characteristics suitable for the ensemble
and not imposed by the assembly are surprising and indicative of an active two-way selection pressure from the ground up
as well as from the top down \cite{Iino}. Initial assumptions, that the rotary catalysis mechanism of
ATP synthase is governed by the rotation of the central $\gamma\delta\epsilon$ unit  that forces each SUB  
into different conformations, ignores these intrinsic propensities that help explain the observation of rotary catalysisF
in rotorless hexamers.

\section{Conclusion}

We examined the reasons why
two proteins with nearly identical topology and  fold but with different primary sequences behave very differently
within the same assembly: one as  a structural protein the other as an enzyme. 
Without distorting the PDB coordinates, we computed the distinct symmetry axes of each protein via
PDB-NMA and discovered unique signatures that help explain their differing behaviors. PDB-NMA
consistently demonstrates the intrinsic flexibility around active sites in enzymes where reactants and
reaction products enter and exit. Structural proteins like SUA do not develop the flexibility characteristics associated
with interdomain motilities,  where similarly disposed regions move, more simply, with block-like character.
Active site motility is enabled by relatively low NBI or packing densities in critical junctures that create sufficient
space to permit  opening and closing movements. Any resultant structural weakness is offset by local
stiffness elements, including  $\beta$ sheets, helices and prolines.

In addition to developing insights into the particular three-dimensional architecture of a stably folded protein,
PDB-NMA demonstrates how locally stiff regions may obtain very high temperature factors. Residues with high
experimental Debye-Waller factors likely are not disordered, but orderedly mobile, an observation supported
not only by crystallographic data but also by MD simulations that observe similar long term behavior as NMA.
Enzymes function as machines that are firm and flexible, and with flexibility characteristics that are reliable and reproducible.

We deformed
the coordinates of a closed $\beta$ chain along the directions of its slowest three modes to achieve better overlap
with the coordinates of the open  chain. An initial RMSD of 4\AA~ between these structures was thereby reduced
to 2\AA, without creating steric clashes. The resultant structure achieves close inter-domain fits of the head, foot and torso 
and less close fits for intra-domain shifts, especially of the nucleotide binding region. During the structural
transformation of the $\beta$ subunit  this region undergoes major shifts in solvation, a feature not modeled by the current
analysis and likely a major reason for the poorer fit. Finally,
PDB-NMA permits examination of observed and computed temperature factors to search for clues in mobility
profiles of different chains. In the case of the three $\alpha$ subunits in 2JDI, two distinct B-factor profiles were
observed that might be indicative of modal suppression of the slowest mode in these two chains.

\acknowledgements

I thank Prof. Garc{\'i}a-Hern{\'a}ndez for providing  data of the MD simulation of the liganded and unliganded
$\beta$ subunits of PDB entry 2JDI and Daniel Ben-Avraham for many helpful discussions.
This work is supported in part by the M. Hildred Blewett Fellowship of the American Physical
Society, www.aps.org. 

\bibliography{myrefs}

\end{document}